%% file: main.tex
\documentclass[conference]{IEEEtran}
\usepackage{cite}
\usepackage{amsmath,amssymb,amsfonts}
\usepackage{algorithmic}
\usepackage{graphicx}
\usepackage{textcomp}
\usepackage{xcolor}
\usepackage[hyphens]{url}
\usepackage{siunitx}
\usepackage{subfig}

\newcommand{\STTmED}{0.2889}
\newcommand{\STTmER}{0.0185}
\newcommand{\STTmEB}{0.0187}
\newcommand{\STTmLD}{0.0006716}
\newcommand{\STTmLR}{0.0001209}
\newcommand{\STTfED}{0.0040}
\newcommand{\STTfER}{0.0002}
\newcommand{\STTfEB}{0.0147}
\newcommand{\STTfLD}{0.0000028}
\newcommand{\STTfLR}{0.0000006}
\newcommand{\fasterMNIST}{3.36}
\newcommand{\fasterHAR}{4.28}
\newcommand{\fasterADULT}{11.29}
\newcommand{\EnergyMNIST}{1,188,716}
\newcommand{\EnergyHAR}{851,282}
\newcommand{\EnergyADULT}{31,736}
\newcommand{\name}{RODENT}

\def\BibTeX{{\rm B\kern-.05em{\sc i\kern-.025em b}\kern-.08em
    T\kern-.1667em\lower.7ex\hbox{E}\kern-.125emX}}

\title{Towards Homomorphic Inference Beyond the Edge}
\author{Salonik Resch\\
resc0059@umn.edu
\and
Zamshed I. Chowdhury\\
chowh005@umn.edu
\and
Husrev Cilasun\\
cilas001@umn.edu
\and
Masoud Zabihi\\
zabih003@umn.edu
\and
Zhengyang Zhao\\
zhaox526@umn.edu
\and

\\
\qquad \quad Jian-Ping Wang\\
\qquad \quad jpwang@umn.edu
\and 
\\
Sachin Sapatnekar\\
sachin@umn.edu
\and
\\
Ulya R. Karpuzcu\\
ukarpuzc@umn.edu

}
\begin{document}
\maketitle
\thispagestyle{plain}
\pagestyle{plain}

%%%%%% -- PAPER CONTENT STARTS-- %%%%%%%%

\begin{abstract}
Beyond edge devices can function off the power grid and without batteries, enabling them to operate in difficult to access regions. However, energy costly long-distance  communication required for reporting results or offloading computation becomes a limitation. Here, we reduce this overhead by developing a beyond edge device which can effectively act as a nearby server to offload computation. For security reasons, this device must operate on encrypted data, which incurs a  high overhead. We use energy-efficient and intermittent-safe in-memory computation to enable this encrypted computation, allowing it to provide a speedup for beyond edge applications within a power budget of a few milliWatts.
\end{abstract}
\input{sections/introduction}
\input{sections/problemStatement}

\section{Background}
\input{sections/homomorphic}
\input{sections/svm}
\input{sections/mtjs}

\input{sections/architecture}
\input{sections/evaluation}
\input{sections/limitations}
\input{sections/related}

\input{sections/conclusion}

%%%%%%%%% -- BIB STYLE AND FILE -- %%%%%%%%
\bibliographystyle{IEEEtranS}
\bibliography{refs}
%%%%%%%%%%%%%%%%%%%%%%%%%%%%%%%%%%%%

\end{document}

%% file: sections/introduction.tex
\section{Introduction}
Devices are now being built which can operate without batteries or other constant power sources \cite{gobieski2019intelligence}. This allows these devices to be deployed in a wide variety of environments, where traditional power sources are typically unavailable. Such environments are referred to as \emph{beyond the edge}, and can reside inside the structures of buildings \cite{manic2016intelligent}, in the remote wilderness \cite{gobieski2019intelligence}, in outer space \cite{lucia2021computational}, inside the human body \cite{greenspan2016guest}, or simply scattered throughout a city. The extreme deployment capability opens up many new applications.

A significant limitation of these devices, however, is the very strict power budget.
%potentially strict power limitations. 
Available power sources 
%of energy 
range from less than \SI{1}{\micro \watt} from RF energy \cite{kim2014ambient,yildiz2009potential,pinuela2013ambient}, to \SI{60}{ \micro \watt} from thermal energy harvesters \cite{mahan1997thermoelectric,kim2014ambient}, up to \SI{100}{\milli \watt} per square centimeter from solar panels~\cite{danesh2011photovoltaic,collado2013conformal,green2018solar,jaffe2013energy,kim2014ambient}. 
%which can provide  
Depending on the specific deployment location, devices could be limited to only the weakest power sources. For example, if embedded within the walls of a building, a device will have no access to sunlight and will have to rely on either RF or thermal energy. Hence, beyond edge devices have to operate effectively while consuming minimal power. Energy efficiency determines the  performance 
%will be determined mostly by  
\cite{gobieski2019intelligence}, as typically most time is
%will be 
spent waiting for sufficient energy. 

A potential solution to this power limitation is to perform the computation in a remote server, rather than on the device itself. Beyond edge devices can collect sensor data and then send it to a server to be processed, rather than performing computation locally. This can provide a benefit if intense processing is required. \emph{Homomorphic encryption} can enable the use of untrusted servers, where the device can encrypt data and the server can process the data without decrypting it \cite{van2021practical}. However, this data transmission itself can be costly. If the server is a few \SI{}{\kilo \meter} away, the cost can be as much as \SI{400}{ \micro \joule} per bit \cite{bouguera2018energy}. To send unencrypted data, for example a single sample of the MNIST data set \cite{MNIST} would consume \SI{2.5}{\joule}. For encrypted data, for most practical purposes, the smallest homomorphic encryption configuration will consume approximately 400,000 bits \cite{sealcrypto}. In this case, transmission would cost \SI{160}{\joule}, well beyond what is practically obtainable by beyond edge devices.  

However, the transmission cost can be significantly reduced if it only needs to travel a short distance. For example, within tens of meters, Bluetooth Low Energy (BLE) can consume as little as \SI{158}{ \pico \joule} per bit \cite{rosenthal2019158}. Hence, if the server is extremely close by, offloading computation is still tractable. Still, it is clearly impossible to deploy a typical server within tens of meters of all beyond edge devices. Hence, in this work, we propose a beyond edge accelerator which can serve as a local ``mini-server''. If this accelerator can be deployed in a location which has relatively more power available, such as on a rooftop where sunlight can be collected, it can provide an effective means for computational offload. 

While this beyond edge accelerator is a trusted device, it introduces new security concerns. As it is deployed beyond edge, in potentially insecure environments, the device itself may be compromised or stolen. In order to maintain security, the accelerator must store and operate on only encrypted data. Otherwise, sensitive information (such as machine learning models), could be compromised. Hence, the accelerator must perform \emph{homomorphic} computation, which allows data to remain encrypted for the entire process. Unfortunately, this introduces a significant challenge, as homomorphic computation is orders of magnitude slower and more energy costly than standard computation. 
In this paper we investigate how 
%However, we are able 
to avoid much of this overhead with two approaches. 

First, we perform only linear operations on the accelerator, such as multiplication and addition, which have a much lower homomorphic overhead (leaving the remaining non-linear operations to be performed on the original device).  If most required operations are linear, such as in \emph{support vector machine} (SVM) inference used in this work, this process remains efficient. Second, we are able to keep the computational depth low, which minimizes the amount of noise. This allows us to avoid the process of \emph{bootstrapping}, which is the most costly component of homomorphic computation. Despite these advantages
, homomorphic computation remains highly energy costly. Consequently, the accelerator must be extremely energy efficient in order to compensate. 

An additional challenge of beyond edge deployment is that the accelerator must also be able to tolerate power outages. Checkpointing and guaranteeing correctness has significant overhead for energy efficiency and complexity \cite{lucia2017intermittent}. We design the accelerator to make use of non-volatile processing-in-memory (PIM), which not only provides high energy efficiency, but inherent resilience to interruptions and low-cost checkpointing mechanisms \cite{resch2020mouse}.

The remainder of the paper is structured as followed. In Section \ref{sec:problemStatement} we define the specific problem  this paper addresses and discuss the requirements for a satisfactory solution. We briefly describe homomorphic computation in Section \ref{sec:homomorphic} and SVMs in Section \ref{sec:svm}. We introduce the memory devices used in the accelerator and describe how we can perform logic with them in Section \ref{sec:mtjs}. The architecture of the accelerator and its operating semantics are covered in Section \ref{sec:arch}. The evaluation is contained in Section \ref{sec:evaluation} and a discussion of the limitations of our work is in Section \ref{sec:limitations}. Related work is covered in Section \ref{sec:related} and finally we conclude in Section \ref{sec:conclusion}.

%% file: sections/problemStatement.tex
\section{Problem Statement and Definitions}
\label{sec:problemStatement}
In this section we cover the basic scenario covered and the assumptions made about the problem. We assume that there are three types of devices in play. The FLY \emph{(in the wall)}, the \name\ \emph{(on the roof)}, and the remote TURTLE.

The TURTLE is a distant and secure server in a secure environment with no power restrictions. It is the intended destination for all final results of interest. 

The FLY is a beyond edge device running on harvested energy. It acquires data with sensors, with the intention of reporting interesting results back to the TURTLE. The FLY is in an concealed environment, where the device itself is secure. Hence, it can operate on secure (un-encrypted) data and has access to a private key. Additionally, due to being in a highly concealed environment (eg. inside a wall), it is limited to RF or thermal energy, which is in the microwatt range \cite{kim2014ambient}. Hence, it must operate under a very low power budget.

The \name\ also operates beyond the edge and relies on harvested energy. However, it is in a less concealed environment and is insecure, the device itself may taken be by an not trusted entity. An advantage of being in a less concealed environment (e.g., on a rooftop), is that is has a higher power budget. For example, it may be able to harvest sunlight, which can provide a power budget of tens of milliwatts \cite{kim2014ambient}. Despite being insecure, \name\ can make use of sensitive information (such as proprietary machine learning models) to process information without introducing security leaks by using encryption. All sensitive data remains encrypted on \name\ at all times. \emph{Homomorphic} computing \cite{BFVbrakerski2012fully,BFVfan2012somewhat}, covered in Section \ref{sec:homomorphic}, enables computation on this encrypted data.

\begin{figure}
    \centering
    \includegraphics[scale=0.35]{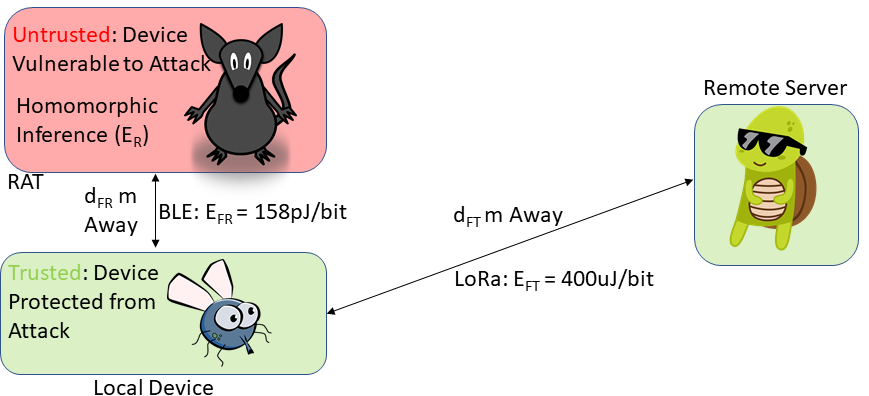}
    \caption{The FLY can transmit data to nearby \name\ to offload computation. Final destination of results is the distant TURTLE.}
    \label{fig:problemStatement}
\end{figure}

The relative placements of the devices are shown in Figure \ref{fig:problemStatement}. TURTLE is $d_{FT}$ meters from FLY, and it takes $E_{FT}$ \SI{}{\micro \joule} per bit to transmit from FLY to TURTLE. \name\ is $d_{FR}$ meters away from FLY and it takes $E_{FR}$ \SI{}{\micro \joule} per bit to transmit from FLY to \name. In this scenario, $d_{FT} >> d_{FR}$ because \name\ can be deployed into similar environments as FLY. Hence, FLY can much more easily communicate with \name\ than with TURTLE.

If the FLY wishes to report back to the TURTLE, it has 3 options. 
\begin{enumerate}
    \item [(1)] Send all sensor readings to the distant TURTLE. TURTLE will be in charge of all computation.
    \item [(2)] Perform processing (inference) locally and only send TURTLE the meaningful results.
    \item [(3)] Offload processing (inference) to the nearby \name\ and then only send the meaningful results to TURTLE.
\end{enumerate}

Option 1 is the simplest, however, it is also typically inefficient. Communication has a higher energy cost per bit than computation. Meaningful sensor data (capturing relevant information) is rare, hence transmitting all data is wasteful \cite{gobieski2019intelligence}. Assume that (non-homomorphic) inference on FLY takes $E_F$ \SI{}{\micro \joule} and the probability that a sample contains results of interest is $\alpha$. Then, performing inference locally is more efficient if
\begin{equation}
    E_F + \alpha \times E_{FT} <
    E_{FT}
    \label{eq:localInference}
\end{equation}

This equation almost always holds as long distance communication (beyond 3km) with low-power long-range (LoRa) hardware is relatively energy costly, consuming up to \SI{400}{ \micro \joule} per bit \cite{bouguera2018energy}. For example, sending data samples (such as an MNIST image sample\cite{MNIST}) would consume $E_{FT} =$ \SI{2.5}{\joule} per sample. Alternatively, the energy required for local inference (with standard beyond edge hardware) is in the range of $E_F =$\SI{27}{ \milli \joule} \cite{gobieski2019intelligence}. In this case, local inference will be more efficient if $\alpha < 98.9\%$. Given results of interest are relatively rare, this is likely to be the case.

Option 3 replaces local inference (consuming $E_F$ \SI{}{\micro \joule}) on FLY with 

\begin{enumerate}
    \item[(1)] (Optional) Encoding and encrypting sensor data (consuming $E_{encrypt}$) 
    \item[(2)] Transmitting data to \name
    \item[(3)] Homomorphic inference on \name\ (consuming $E_R$)
    \item[(4)] Transmission of results back to FLY (consuming $E_{RF}$)
    \item[(5)]Decrypting results on FLY (consuming $E_{decrypt}$)
\end{enumerate}

For option 3 to be superior to option 2, the following equation must hold:
\begin{equation}
\label{eq:option3}
    E_{encrypt} + E_{FR} + E_{RF} + E_{decrypt} <
    E_F
\end{equation}

In words, it must be more efficient for FLY to transmit to \name\ and receive the results than it is to process the data itself. If the input data collected by FLY is considered sensitive, FLY must encrypt the data before transmission, in order to maintain security in un-trusted environments. $E_{encrypt}$ and $E_{decrypt}$ can be done with \SI{0.06}{ \milli \joule} (using the same parameters as this paper) with specialized hardware for homomorphic encryption on edge devices \cite{van2021practical}. If the inputs are not sensitive, then they can be sent as raw data and $E_{encrypt}$ can be dropped from Equation \ref{eq:option3}. However, \name\ will still return encrypted data in order to protect its ML model, hence $E_{decrypt}$ is still required. Due to the close proximity of FLY and \name\ (in the order of meters), communication cost is significantly reduced. Technologies such as Bluetooth Low Energy (BLE) can be used, with some configurations offering as little as \SI{158}{ \pico \joule} per bit \cite{rosenthal2019158}. At such efficiency, (un)encrypted data can be sent to \name\ ($E_{FR}$) and encrypted results can be transmitted from \name\ to FLY ($E_{RF}$) with less than (\SI{1}{\micro \joule}) \SI{80}{ \micro \joule}. Hence, BLE can provide communication which will cost FLY less energy than local processing. FLY will require both BLE (for communication with \name) and LoRa (for communication with the distant TURTLE).

The other condition required for option 3 to be viable is that the homomorphic computation energy on \name\ ($E_R$) must be reasonable. While operating beyond edge, \name\ has a higher energy budget than FLY. However, homomorphic computing has a very large overhead that can easily become excessive. \name\ must be sufficiently efficient in order to return results to FLY promptly. For \name\ to provide any benefit, the following equation must hold:
\begin{equation}
    \frac{E_{FR}}{P_F} + \frac{E_R}{P_R} + \frac{E_{RF}}{P_R} < 
    \frac{E_F}{P_F}
\end{equation}
where $P_F$ is the power available to FLY and $P_R$ is the power available to \name. In words, \name\ must have sufficient efficiency to homomorphically compute and return data within its power budget faster than FLY can non-homomorphically compute within its power budget.
In this paper, we design \name\ to perform such computations within a reasonable energy budget.

%\subsection{Related Problem Statement}
%The previous problem statement is similar to that analyzed by van der Hagen et. al. \cite{van2021practical}. In their problem statement, a secure beyond edge device offloads encrypted data to a remote untrusted server for processing. The large amount of homomorphic computation perform on the server (which has no power restrictions) justifies the transmission cost. In contrast, in our problem statement, \name\ acts as a nearby ``mini-server'', where it can reduce the need for long-distance communication. Hence, not only is the sensor device ``beyond the edge'', but so is the server. However, this means \name\ must maintain a low power budget and remaining resilient to intermittent operation.

%% file: sections/homomorphic.tex
\subsection{Homomorphic Computing}
\label{sec:homomorphic}
Homomorphic encryption allows computation to occur on encrypted data, typically with a significant time and energy overhead \cite{BFVbrakerski2012fully,BFVfan2012somewhat}. However, some operations are more efficient than others. For example, linear operations tend to have the least overhead. The vast majority of the operations performed in SVMs are linear multiplications and additions, explained in Section \ref{sec:svm}. We choose to perform only these linear operations on \name, leaving the final, non-linear operations to be performed after decryption on FLY.

We use the BFV scheme \cite{BFVbrakerski2012fully,BFVfan2012somewhat}, which provides integer arithmetic and is available on both Microsoft SEAL \cite{sealcrypto} and PALISADE \cite{polyakov2017palisade}. First, a vector of data is \emph{encoded} into a \emph{plaintext}, which is a set of coefficients for a polynomial. Then, the plaintext is \emph{encrypted} into a \emph{ciphertext}. The length of the input vector is equal to the number of coefficients in both the plaintext and ciphertext (also called the degree of polynomial modulus). The length of the vector sets a noise budget, where larger vectors have larger noise budgets. All homomorphic computations consume some of the noise budget, and if the budget is exceeded, the data can no longer be successfully decrypted. Testing with Microsoft SEAL \cite{sealcrypto}, we found 4,096 was the minimum vector size that could successfully complete the SVM computations. The \emph{coefficient modulus} is the maximum value of the ciphertext coefficients, which is represented in a residue number system consisting of three 36-bit prime numbers. A ciphertext consists of two such polynomials. Hence, the total memory required by our ciphertexts is 

\begin{equation}
    2 \times 3 \times 36 \times 4096 = 110 KB
\end{equation}

Homomorphic addition is relatively straightforward, consisting of element-wise additions of the vectors involved. Modulus operations must be performed routinely to prevent the coefficients from growing excessively large. Given that the coefficient modulus is known ahead of time, modulus can be implemented with efficient (virtual) shift, add, and subtract operations in the memory \cite{nejatollahi2020cryptopim}.

Homomorphic multiplication is significantly more complicated and consists of multiple steps. The primary components are the number theoretic transform (NTT), which is an integer variant of the FFT, and scales and base conversions \cite{turan2020heaws}. We follow the computation steps laid out by {\"O}zerk et. al. \cite{ozerk2021efficient} for NTT and by Al Badawi et. al. \cite{al2019implementation} for all other steps.

%% file: sections/svm.tex
\subsection{Support Vector Machines}
\label{sec:svm}
Support vector machines (SVM) are popular machine learning (ML) algorithms which perform well on small problems. The advantages of SVMs are a strong theoretical foundation and proven performance on a wide variety of problems \cite{noble2006support}. SVMs are trained by inspecting input samples to a problem, where samples which are indicative of each classification are identified. After training is complete, these samples are referred to as \emph{support vectors}, and are used to compare to new inputs. The new inputs are classified based on how similar they are to support vectors from each class. By construction, a standard SVM model produces a binary classification, producing a classification of $\pm 1$. We use a simple extension to enable multi-class classification, where a separate SVM is trained to identify each class (the one-versus-all method \cite{duan2005best}). For example, to classify MNIST, 10 SVMs are used where each identifies one digit 0-9.

A specific advantage for SVMs is that they consist mostly of linear operations, which can be implemented relatively efficiently in homomorphic computing. This provides an advantage over neural networks, for which sophisticated strategies must be invoked to reduce the overhead of frequent non-linear operations \cite{reagen2021cheetah,chao2019carenets}. SVM inference involves finding the dot product of the input sample with each of the support vectors. This involves purely multiplications and additions. At the very end, the dot products are squared (multiplication by self), summed, and finally compared, which represents a non-linear operation. While complete SVM training and inference has been successfully demonstrated using homomorphic operations \cite{park2020he}, we opt to perform only the initial multiplication and summations on \name. We leave the final squaring and sum to be performed on FLY. This prevents us from exceeding the computational depth allowed by our homomorphic encryption (without performing bootstrapping) and avoids the high overhead incurred by the final sum (which requires highly expensive rotation operations to sum elements in the same ciphertext \cite{reagen2021cheetah}).

%% file: sections/mtjs.tex
\subsection{(Digital) Computing with Magnetic Tunnel Junctions}
%and Basic Logic}
\label{sec:mtjs}
Magnetic Tunnel Junctions (MTJ) are resisistive memory devices consisting of two magnetic layers, a \emph{free layer} and a \emph{fixed layer}. The polarity of the free layer can change but the fixed cannot. When the magnetic layers are (not) aligned, the MTJ is in the  parallel $R_P$  (anti-parallel $R_{AP}$) state and has a low (high) resistance. Passing a sufficient amount of current ($I_{switch}$) through the MTJ from the free (fixed) layer to the fixed (free) layer will set the MTJ into the $R_P$ ($R_{AP}$) state.

\begin{figure}
    \centering
    \includegraphics[scale=0.25]{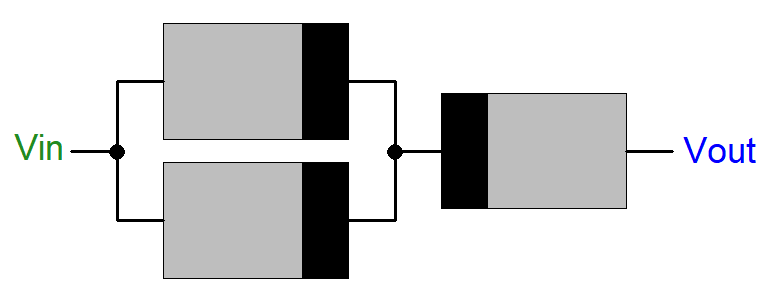}
    \caption{Circuit for logic with MTJs. Input MTJs (in parallel) are connected in series with an output MTJ. A gate-specific voltage applied across $V_{in}$ and $V_{out}$ drives a current which will switch the output MTJ depending on the state of the input MTJs, following the truth table of the corresponding gate. Fixed (free) layer of MTJ is shown in light (dark).  }
    \label{fig:logic}
\end{figure}

Logic can be performed with MTJs via thresholding \cite{chowdhury2017efficient,louis2019performingMTJMAGIC}, where input MTJs (in parallel) are in series with an output MTJ, as shown in Figure \ref{fig:logic}. For example, a NAND can be peformed by presetting the output MTJ to logic 0 ($R_P$). A (gate-specific) voltage is applied such that electrons flow from the inputs to the output. If both input MTJs are logic 1 ($R_{AP}$, high resistance) the current through the output will be \emph{less} than $I_{switch}$, and it will remain 0. However, if \emph{either} input MTJ is logic 0 ($R_P$, low resistance) the current will be \emph{greater} than $I_{switch}$ and the output will switch to 1. In a nutshell, under the fixed gate-specific voltage the current through the output changes as a function of the input states, and only incurs switching (a change in the output state from the preset) according to the truth table of the corresponding gate. Different logic operations, including NOT, AND, and (N)OR can be performed in identical fashion, with different output preset values and %input 
gate-specific voltages, respectively.

%% file: sections/architecture.tex
\section{Architecture Design}
\label{sec:arch}
\begin{figure}
    \centering
    \includegraphics[width=.48\textwidth]{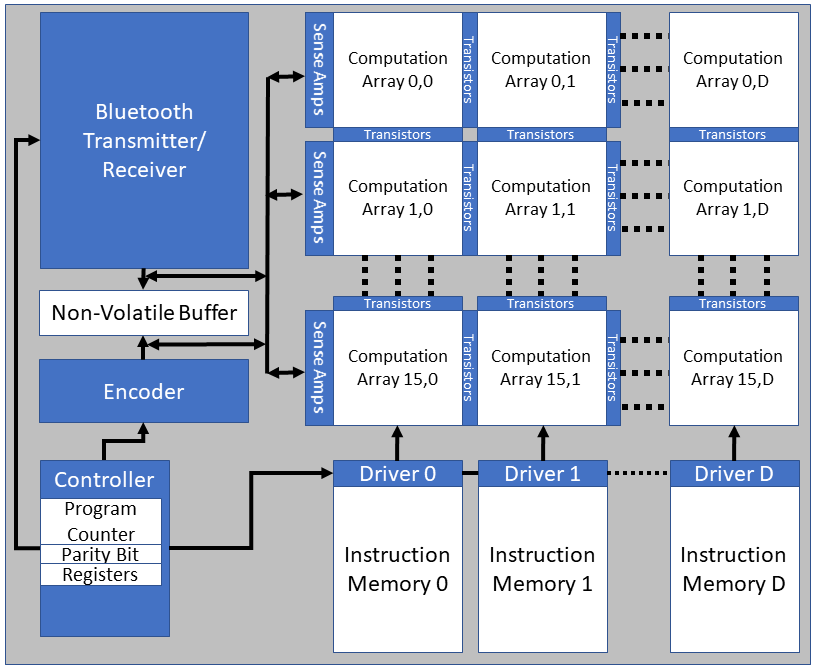}
    \caption{\name\ architecture containing non-volatile computation arrays and instruction memory along with volatile circuitry for driving operations. Only the first column of computation arrays has sense amplifiers, allowing the peripheral circuitry to both read and write from them. Data transfer for all other computation arrays occurs with logic operations. As proposed by Gupta et. al.\cite{gupta2018felix} transistors between neighboring arrays enables them to participate in the same logic operation.}
    \label{fig:architecture}
\end{figure}

\name\ must perform homomorphic computation while both remaining within a low power budget and being resilient to intermittent operation. We design an accelerator using a non-volatile processing-in-memory (PIM) substrate
%based 
%logic 
which is shown to be both highly energy efficient and inherently intermittent resistant \cite{resch2020mouse}. We augment the PIM capability with simple, intermittent safe protocols and hardware for reception, transmission, and encoding. The architecture is shown in Figure \ref{fig:architecture}. First, in Section \ref{sec:archHardware} we describe the hardware contained within \name, then in Section \ref{sec:archSemantics} we cover how different system components interact with each other, and tolerate intermittent operation. 

\subsection{Hardware}
\label{sec:archHardware}

\subsubsection{Computation Arrays}
\label{sec:hardwareComputationArrays}
\begin{figure}
    \centering
    \includegraphics[width=.4\textwidth]{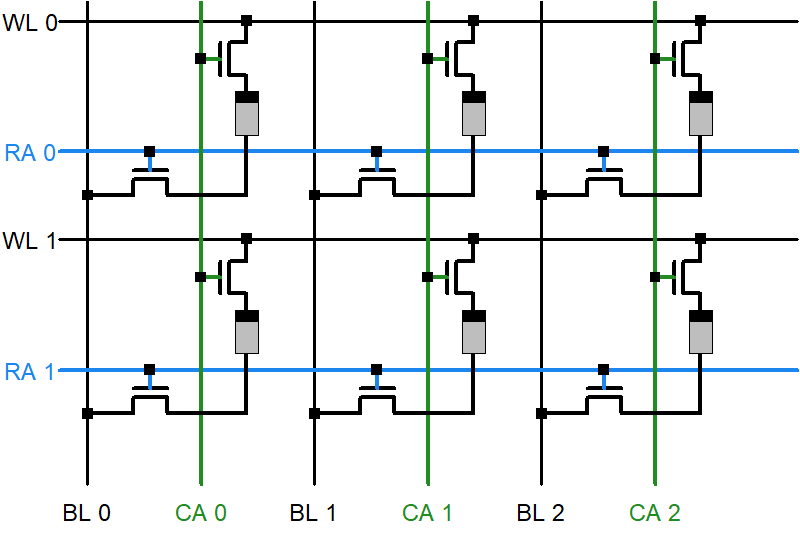}
    \caption{The 2T-1M cross-bar array architecture. Logic operations can be driven in memory by applying voltages either along bitlines (BL) or wordlines (WL). The access transistors, activated by the signal column activate (CA) and row activate (RA), remove sneak paths.}
    \label{fig:array}
\end{figure}

\begin{figure}
    \centering
    \subfloat[Row Logic]{\includegraphics[width=.2\textwidth]{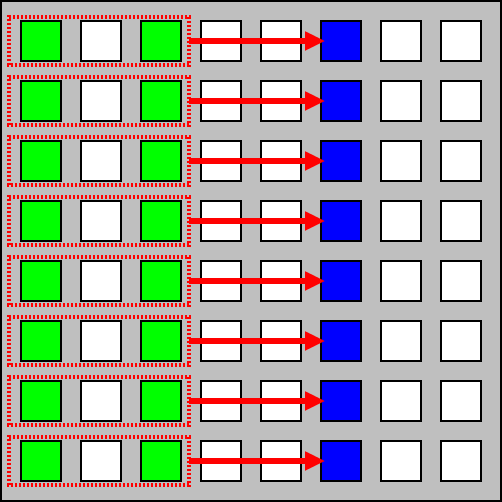}}\quad
    \subfloat[Column Logic]{\includegraphics[width=.2\textwidth]{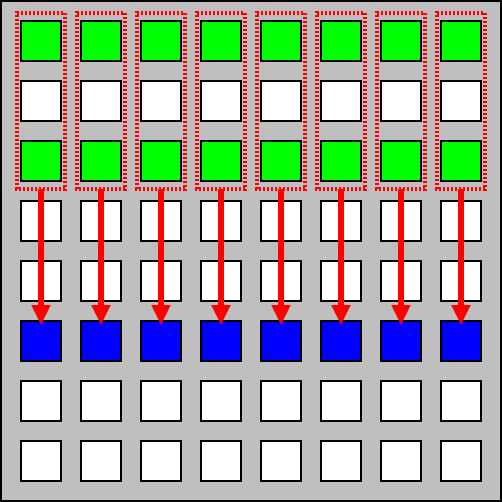}}
    \caption{Computation arrays can perform row logic or column logic. Inputs are in green/light and outputs are in blue/dark.}
    \label{fig:parallelism}
\end{figure}

\begin{figure}
    \centering
    \includegraphics[width=.4\textwidth]{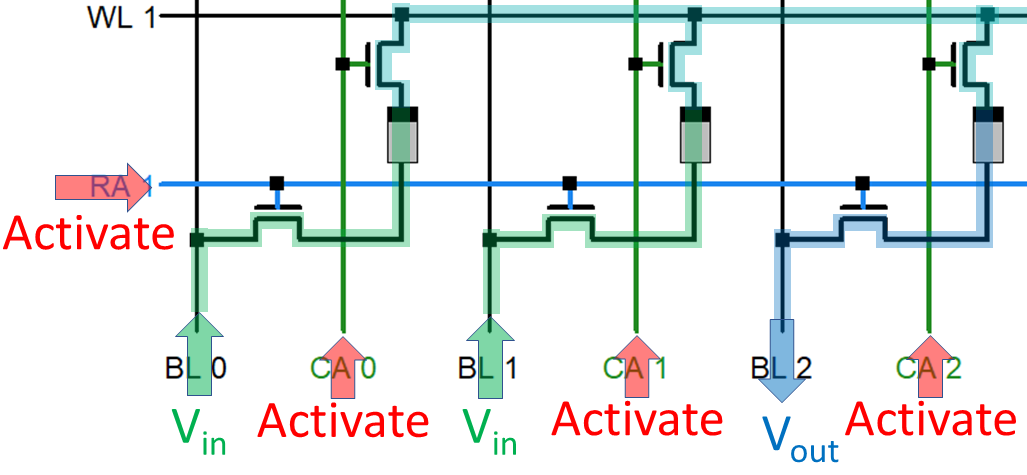}
    \caption{Voltages $V_{in}$ and $V_{out}$ are applied to the bitlines to drive logic operations. Row Activate (RA) and Column Activate (CA) set on which rows and columns the operations occur. Shown is a row-logic operation, which implements the same functionality as the circuit shown in Figure \ref{fig:logic}. }
    \label{fig:lim}
\end{figure}

The primary component of \name\ are arrays of MTJ devices used for computation. We call these \emph{computation arrays}. These arrays both hold data and perform all homomorphic computation on it. The array architecture and cell design is shown in Figure \ref{fig:array}, which uses two transistors. This architecture allows for the logic operations discussed in Section \ref{sec:mtjs} to occur in either the rows or columns of the memory array by applying voltages along the bitlines (BL) or the wordlines (WL). The access transistors, controlled by row activate (RA) and column activate (CA), remove potential sneak paths from the array. An example of row-logic is shown in Figure \ref{fig:lim}, where voltages $V_{in}$ and $V_{out}$ are applied to the bitlines, implementing the logic circuit shown in Figure \ref{fig:logic}. Many logic operations can be performed in parallel in each row (column) simultaneously, as long as the inputs and output reside in the same columns (rows) as shown in Figure \ref{fig:parallelism}. Having both row- and column-wise logic allows data to be moved efficiently within each array, removing the need for energy costly read and write operations. This computational capability is equivalent to digital cross-bar arrays \cite{talati2016logicMAGIC,bhattacharjee2020contra}, but without the sneak paths which waste energy and introduce correctness concerns \cite{li2021memristiveSneakSurvey}. In row- (column-) logic, for each operation the column (row) addresses of the input(s) and output must be specified, along with which rows (columns) are participating in the operation.  When rows or columns are performing computation we refer to them as \emph{active}.

 Using two transistors per cell significantly increases the area per cell from \SI{0.0012}{ \micro \meter}$^2$ \cite{mondal2019situ} to \SI{0.03815}{\micro \meter}$^2$ \cite{zabihi2020analyzing}. However, two transistors are essential for removing cross-bar sneak paths \cite{li2021memristive}, which would significantly increase energy consumption. As energy efficiency is paramount for beyond edge devices \cite{gobieski2019manic}, 
 this cost in area is necessary. 
 %the cost in area is necessary. 
 Due to the very large amount of data required by homomorphic computing, the area overhead for computation arrays can become substantial. For the smallest benchmark, ADULT, the computation arrays consume \SI{6.72}{ \milli \meter}$^2$. For the largest benchmark, MNIST, the area is \SI{377}{ \milli \meter}$^2$. For reference, the TI-MSP430FR5994, commonly used as a sub-component of beyond edge systems \cite{gobieski2019intelligence}, consumes roughly \SI{100}{ \milli \meter}$^2$.

 As described in Section \ref{sec:homomorphic}, we use ciphertexts with 4,096 elements. Hence, it is ideal to have 4,096 rows in order to store all elements and enable parallel element-wise homomorphic multiplications and additions. However, due to parasitic bitline/rowline resistance and capacitance, arrays are limited to 1024x1024 \cite{zabihi2020analyzing}. Hence, we use 16 rows of computation arrays, where each array is 512x512. Similar to Gupta et. al. \cite{gupta2018felix}, a row of transistors are used to conditionally connect to the bitlines of neighboring computation arrays. This allows logic operations to transfer bits between the neighboring arrays. 

Each row of the computation arrays must store all required data for its given computation. The most space intensive subroutine of the homomorphic SVM is the initial multiplication, which can be between two ciphertexts or between a ciphertext and a plaintext. A single element of each polynomial is assigned to each row. Hence, each row requires $2 \times 2\times 3 \times 36 = 432$ bits for the ciphertexts, 36 bits for \emph{twiddle factors} for NTT operations \cite{ozerk2021efficient}, plus additional bits for temporary workspace. This easily fits within two columns of computation arrays (1024 bits). However, a third column of computation arrays is required to store additional twiddle factors required for different stages of the NTT algorithm. In total, $log_2 (4096) = 12$ twiddle factors are required in each row.

Since logic operations can be used to transfer data within and between computation arrays, only the cells which store the final results (which are sent to the Bluetooth transmitter) need to be read. Hence, the vast majority of the computation arrays do not require sense amplifiers. As shown in Figure \ref{fig:architecture}, only the first column of computation arrays contain sense amplifiers, which allows them to be used as a non-volatile buffer for the BLE transmitter/receiver and the encoder. Hence, input and output is passed through the first column on computation arrays.

\begin{figure}
    \centering
    \includegraphics[width=.25\textwidth]{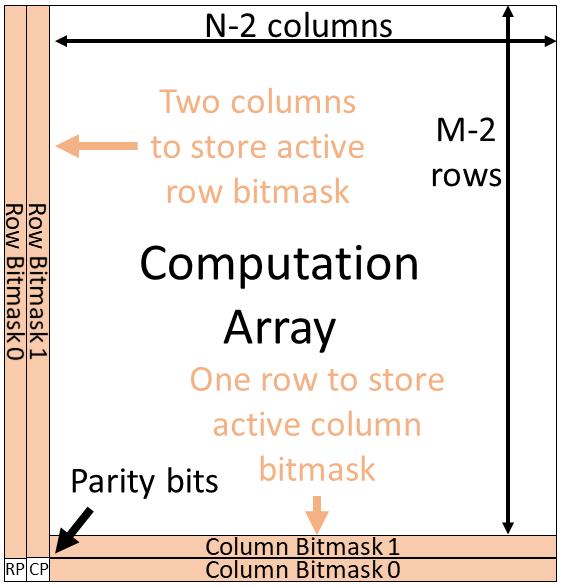}
    \caption{Two rows and columns of each computation array is dedicated to hold  bitmasks for the active rows and columns. Standard logic and write operations can set the bitmask and dedicated instructions apply the bitmask to the peripheral circuitry. Two bitmasks are required for each in order to prevent corruption when modifying the bitmask. Parity bits, row parity (rp) and column parity (cp) indicate which bitmask is valid.}
    \label{fig:registers}
\end{figure}

In addition to the memory data, computation arrays must also maintain which rows or columns are currently active. For this purpose, we could use two dedicated non-volatile registers in each computation array - with as many bits in the registers as there are rows and columns in the computation array. Such registers can act as a bitmask for logic operations. However, rather than creating an additional hardware register, we can embed the registers into the memory itself, by dedicating a single row and column for each register. This allows registers to be written with standard logic and write operations. Dedicated instructions use the registers to activate the peripheral circuitry. For correctness guarantees which will be discussed in Section \ref{sec:operatingComputation}, two copies of the registers are required for both rows and columns, as shown in Figure \ref{fig:registers}.

\subsubsection{Drivers and Instruction Memory}
Signals must be sent into the computation arrays to perform logic gates. For every operation in each array, we need to  specify which logic gate is being performed (what voltage to apply) and the addresses of the input(s) and the output. Just as in prior work \cite{resch2020mouse}, the input and output addresses are specified with each instruction. Additionally, as noted in the previous paragraph, we must known which rows (or columns) are currently \emph{active}. The currently active rows (or columns) are specified by the bitmask column and row within each computation array.  

As each array could perform computation independently, many different gates, inputs, and outputs may need to be specified simultaneously. To efficiently distribute these signals, computation arrays are grouped into columns. Computation arrays in the same column act as a single unit and are driven by the same control signals. For each column of computation arrays, there is an associated \emph{driver} and non-volatile \emph{instruction memory} buffer. 

The instruction memory arrays store the operations that each column of computation arrays is to perform (NOT, (N)AND, (N)OR) along with the row/column indices of the input(s) and the output of every operation.  Each instruction requires the following information. 
\begin{enumerate}
    \item[i.] Opcode specifying the logic operation (3 bits)
    \item[ii.] Up to three addresses for input(s) and output (12 bits each)
    \item[iii.] Whether it is a row or column operation (1 bit)
\end{enumerate}

To be specific, a driver is a CMOS circuitry that initiates the logic within the computation arrays. On command from the controller, each driver reads an instruction from the specified address and decodes it. Then it sends the input and output addresses to the row/column decoders and drives the appropriate voltage along the bit/row lines of the computation arrays. All drivers operate synchronously, executing the instructions at the same address (specified by the controller) in their corresponding instruction memories. 

\subsubsection{Encoder}
\name\ operates on encrypted data in order to keep the ML model secure (encrypted at all times). If the input is also considered sensitive, it should be encrypted on FLY (into a \emph{ciphertext}) prior to transmission to \name, in which case \name\ can immediately begin processing and does not require an encoder. However, if the inputs are not considered sensitive, it will be more efficient to transmit the raw data. In which case the input, once transmitted to \name, must be encoded (into a \emph{plaintext}) in order to properly interact with the encrypted ML model. Homomorphic encoders for edge devices have previously been developed \cite{van2021practical}. We assume \name\ contains such a hardware chip for this purpose. 

\subsubsection{Receiver and Transmitter}
\name\ uses Bluetooth Low Power (BLE) to communicate with the nearby FLY. BLE devices can offer extremely energy efficient communication at short distances \cite{siekkinen2012low,gomez2012overview,kamath2010measuring,sultania2021energy,liu2021comprehensive}. We assume a configuration that enables particularly low power, down to \SI{158}{ \pico \joule} per bit \cite{rosenthal2019158}. BLE will allow communication within a few tens of meters. 

\subsubsection{Controller}
The controller comprises CMOS circuitry which orchestrates the operation of \name. It does not directly orchestrate the operation of the other components, rather it simply turns them off or on, and triggers their operation. It maintains a status register (SR), program counter (PC), activates the BLE transmitter/receiver and encoder, and sends trigger signals to the drivers to perform logic in the computation arrays. 

\subsection{Operating Semantics}
\label{sec:archSemantics}
\begin{figure}
    \centering
    \includegraphics[width=.5\textwidth]{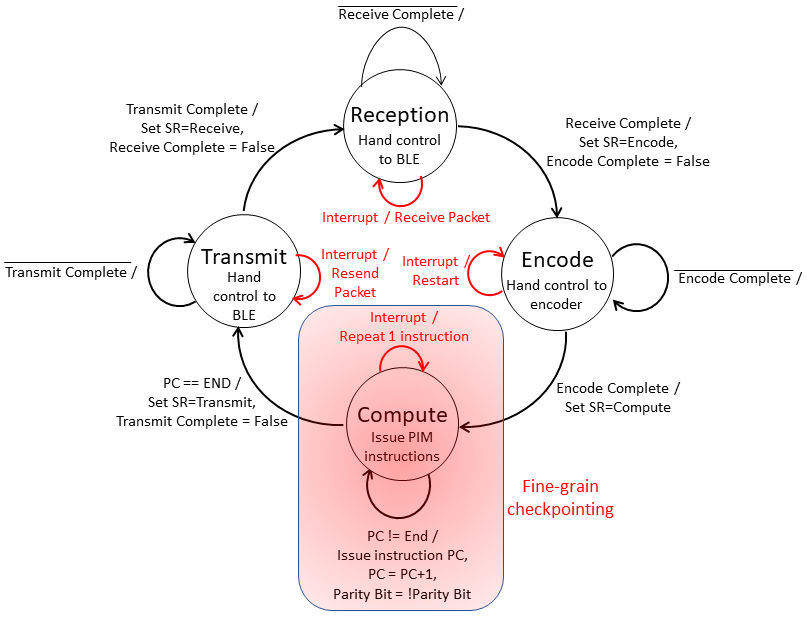}
    \caption{\name\ follows a simple state transition diagram. The controller is responsible for maintaining architectural state variables and ensuring correctness of state transitions. \name\ has efficient checkpointing mechanisms in the compute phase, where it spends the vast majority of the time, due to ideal properties of MTJ devices.  } 
    \label{fig:stateMachine}
\end{figure}

Now that we have described the hardware components of \name, we describe how they work together and how they tolerate intermittency. The state transition diagram is shown in Figure \ref{fig:stateMachine}. \name\ has efficient and fine-grain checkpointing mechanisms in the computation phase, due to the inherent resilience of non-volatile MTJ based memory to power interruptions. \name\ has less efficient checkpointing mechanism in the other phases, where more progress will be lost in the event of a restart.

\subsubsection{Data Reception and Transmission (I/O)}
 Due to intermittent power, \name\ cannot rely on continuous communication. To avoid complexity, we chose for \name\ to operate under a simple protocol. When \name\ is not busy (there is no input being processed), it is open to new input. \name\ will wait for new input from FLY. After fully receiving input data, \name\ enters the encoding phase and will no longer check for or acknowledge new inputs. Only once computation is complete will \name\ re-activate BLE in order to transmit outputs. 
 
 When receiving inputs, the data is stored into a dedicated non-volatile buffer, as shown in Figure \ref{fig:architecture}. The size of the input data is set, so \name\ will know the number of packets it needs to receive. There is a dedicated region of memory for each packet, and there is a \emph{valid bit} for each packet. \name\ will set all valid bits to 0 prior to reception. The valid bit for each packet is set strictly after the packet has been written into memory. If \name\ looses power after the packet is written but prior to setting of the valid bit, the packet is considered invalid and will need to be re-transmitted . Once all valid bits are set (all packets have been received), \name\ will set a \emph{completed} bit, which acts as a signal to \name's controller to move to the next stage.
 
  During reception and transmission, no other components of \name\ are in use. The controller will stall all operation until data transfer is complete.

 \subsubsection{Encoding} 
 \name\ uses specialized hardware accelerators for encoding \cite{van2021practical}. To make the encoding process intermittent safe, we require it to be \emph{atomic}. It reads input from the dedicated non-volatile buffer, performs encoding in full, and then writes output into the first column of computation arrays, shown in Figure \ref{fig:architecture}. Due to being atomic, if it is interrupted at any stage it restarts from the beginning. This guarantees correctness on restart, as none of the inputs have been overwritten, however it can waste energy. There is significant room for optimization (i.e., adding efficient checkpointing mechanisms for the encoder), however, we do not focus on the encoder in this work. The vast majority of time and energy is spent on homomorphic computation, where we optimize the checkpointing process. 
 
 Since the data is written directly into the computation arrays, it will be ready immediately for processing once encoding has completed. Computation arrays can transfer the data where it is required via logic operations during the computation stage. %Likewise, when transmitting output, data is read from this region. Hence, computation arrays can move output data into these computation arrays with logic operations, making them available for reading once completed.
 
 \subsubsection{Computation}
 \label{sec:operatingComputation}
 Computation takes the most time and energy, therefore it is critical to make it resilient to intermittent operation and  energy efficient. Prior work has demonstrated that in-memory logic performed with MTJs is inherently intermittent safe \cite{resch2020mouse}. We exploit this same property to enable high frequency and light-weight checkpointing during computation. However, unlike prior work, we do not need to maintain standard memory functionality. This enables us to remove much energy-inefficient hardware, including most of the sense amplifiers. 
 
 During computation, the controller will broadcast the PC value to all drivers. Each driver reads the corresponding instruction in its instruction memory, decodes it, then drives the appropriate signals to the computation arrays in the same column. A sufficient amount of time is allotted between broadcasts of the PC value such that every instruction is guaranteed to have completed. This process repeats until the program has finished or \name\ runs out of power.
 
 Write and logic operations with MTJs \cite{louis2019performingMTJMAGIC,chowdhury2017efficient,zabihi2018memory} remain correct when interrupted or performed multiple times \cite{resch2020mouse}. This means that if the driver triggers a logic operation in the computation arrays, it can be interrupted 
 \begin{enumerate}
     \item [i.] Before the operation begins
     \item [ii.] In the middle of the operation
     \item [iii.] After the operation has finished
 \end{enumerate}
 and correctness will be maintained. The same logic operation (instruction) can be triggered again, and the output will be as if no interruption had occurred \cite{resch2020mouse}. Hence, a single instruction represents an \emph{idempotent} operation. This means that, after \name\ has restarted, the controller can safely send the same PC value to the drivers and the instruction will be performed (or re-performed) and produce the correct output. 
 
Due to this inherent resilience, correctness of data in the memory is easy to guarantee as long as we progress in program order by only a single instruction at a time. This means we need to checkpoint after every instruction, and not start the next instruction until the previous has been completed. Such frequent checkpointing may be inefficient for more traditional architectures \cite{kim2014ambient,liu2015ambient}, however \name\ can do this with ease. As the computation is occurring in (non-volatile) memory, all data backup happens automatically. All results are permanently stored after every instruction, regardless of the checkpointing strategy. Hence, \name\ can perform frequent checkpointing with low overhead by simply tracking architectural state variables in the controller. For computation, the controller only needs to keep track of the valid PC value. 

As noted in Section \ref{sec:hardwareComputationArrays}, which rows or columns are active is part of the architectural state. A single row and column from each computation array is dedicated to hold the bitmask for active columns and rows. Hence, the active rows and columns are protected by the same mechanism as all other data in memory. Being non-volatile, the bitmasks persist through power interruptions. However, the peripheral circuitry will need to be re-activated on restart. Dedicated instructions, \emph{activate rows} and \emph{activate columns}, restore the peripheral circuitry based on the bitmask values. These instructions are issued whenever the bitmask changes, the instructions switch between row-wise or column-wise, and on re-start.

An advantage of storing the active row/column bitmasks in the computation arrays themselves is that they can be set with standard write and logic operations. A disadvantage is that it introduces potential incorrectness. If a row- (column-) parallel instruction modifies the row (column) bitmask, the instruction relies on the current value of the bitmask. If this operation gets interrupted prior to completion of the instruction, some of the values in the bitmask may have changed. When the instruction is re-performed on startup, the modified value of the bitmask will be used instead. Our solution is to duplicate the bitmasks and maintain parity bits, row parity (RP) and column parity (CP), which indicate which bitmask is valid. Hence, two columns and two rows are dedicated for bitmasks. Instructions are only allowed to modify the currently invalid bitmask, leaving the currently valid bitmask unperturbed. After the invalid bitmask has been successfully modified, the corresponding parity bit can be flipped and the rows or columns re-activated.

\subsubsection{{Controller}}
RAT's controller is responsible for maintaining the state transitions depicted in Figure \ref{fig:stateMachine} and it holds the architectural state variables required for driving computation. It maintains a \emph{status register} (SR), which indicates whether \name\ is in the reception, encoding, computing, or transmission state. Additionally, it maintains the \emph{program counter} (PC) which is sent to the drivers to initiate instructions during the computation state.

Given the relatively little information the controller must maintain, simple checkpointing strategies are sufficient. Both the SR and the PC are duplicated and have an ancillary non-volatile \emph{parity bit}. The parity bit indicates which copy is valid, and it is flipped strictly after each variable is updated. Hence, a currently valid copy of either the SR and PC is never written to, which prevents potential corruption due to an interrupted write operation. Setting the parity bit is an atomic operation, the non-volatile MTJ holding the bit will either be successfully flipped or will remain in its old value (if interrupted during the write operation). If \name\ restarts without flipping the parity bit, the old value of the SR and PC will be used. As explained below, this will waste energy but not introduce incorrectness. If \name\ restarts after flipping the parity bit, effectively all progress has been saved and \name\ will start where it left off. To ensure correctness of these mechanisms, \name\ performs the states and instructions sequentially and strictly \emph{in order}. There is no overlap of instructions, states of the controller, or updates of the architectural state variables. 

During the transmission and reception phases, the controller hands control over to the BLE transmitter/receiver. An efficient protocol for intermittent safe transmission/reception  is beyond the scope of this work -- see Section \ref{sec:limitations} for a discussion.
%on potential improvements. 
However, we note that strategies for noisy transmission with large amounts of packet loss \cite{arabi2018information} could likely be adapted for handling intermittency. \name's controller will hand over control and the simply wait for the \emph{completed} signal, in which case it will transition to the next phase. If interrupted prior to the arrival of the complete signal, \name\ will again hand control over to the transmitter/receiver on restart. In the worst case, an entire data packet will need to be re-transmitted or received. Additionally, if transmission/reception have completed, but the completion signal was not sent prior to shut down,
%. In this case, 
the controller will have to re-check the signal before progressing to the next stage.

The encoding process follows a similar strategy. During the encoding processes there are no checkpoints. Introducing checkpoints can guard against progress loss on restart, however, for this work we assume the encoding process must not be interrupted. Hence, if there is a restart in the encoding state, the controller will instruct the encoder to start from the beginning.

\name\ has efficient and finely-grained checkpointing mechanisms in the compute phase. Due to the ideal properties of MTJ based memory arrays discussed in Section \ref{sec:hardwareComputationArrays}, \name\ can easily checkpoint after every instruction \cite{resch2020mouse}. The controller sends the instruction address to the drivers, which load the instruction and trigger it in the computation arrays. The controller waits a sufficiently long time to guarantee the completion of the instruction, after which it updates the PC and commits the instruction by flipping the non-volatile parity bit. As this checkpoint occurs after each instruction, at most one instruction needs to be re-performed in the case of a power outage. The logic operations performed in the memory are inherently idempotent \cite{resch2020mouse}, (meaning that they can be performed multiple times) and will produce the same result. Hence, the instruction can be interrupted at any stage in its progressing and we can safely restart it when the power is restored. 
 
%While we use the same in-memory logic mechanism as prior works \cite{resch2020mouse,louis2019performingMTJMAGIC,talati2016logicMAGIC}, we modify the array structure to enable a more flexible set of computations. Specifically, we require two-dimensional compute capability 

%% file: sections/evaluation.tex
\section{Evaluation}
\label{sec:evaluation}
We evaluate \name\ performing one homomorphic SVM inference to assess the feasiblity. We can then compare the net impact on FLY, to see if the presence of \name\ provides an improvement in performance. We analyze the scenario where inputs are not sensitive, however, the machine learning models that process them are proprietary, and are therefore sensitive. This means that FLY can transmit non-encrypted sensor data to \name, however \name\ must perform homomorphic inference and return encrypted results to keep the model secure. 

For benchmarks, we use MNIST \cite{MNIST}, Human Activity Recognition \cite{HARrequest,HARweb}, and ADULT \cite{ADULTkohavi1996scaling}. These datasets are representative of the input sizes that will be expected for beyond edge devices. When performing homomorphic inference, the data size is determined by $D$, the dimension of the input sample sizes, and $N$, the length of our homomorphic ciphertexts. The $D$ individual elements from each sample must be in separate ciphertexts, as each element must be summed together (elements within the same ciphertext cannot be added without expensive rotation operations \cite{reagen2021cheetah}). As described in Section \ref{sec:homomorphic}, our ciphertexts are fixed at $N=4096$. The number of support vectors required for each benchmark fill the 4096 element ciphertexts, with the remaining elements padded with dummy data. Hence, regardless of the number of support vectors, each benchmark uses 4096 element ciphertexts. MNIST contains 784 elements (representing a $28 \times 28$ image), HAR contains 561 elements, and ADULT contains 14 elements. For all benchmarks, we normalized the input to fit within 3-bit integers (values 0-7). This lowers transmission cost and prevents arithmetic overflow during computation. We use customized SVM models which use only integer arithmetic and limited the number of support vectors to 4096. Despite these limitations, we were still able to achieve reasonable accuracy relative to full-precision SVMs provided by libSVM \cite{libsvm}, which we accessed through R \cite{R} with the `e1071' package \cite{libSVMR}. The SVM parameters and accuracies are listed in Table \ref{tab:benchmarks}. Using integers allows us to use the BFV homomorphic scheme for integer arithmetic \cite{BFVbrakerski2012fully,BFVfan2012somewhat}, which has a lower overhead. We validated that homomorphic operation produces identical output by using Microsoft SEAL \cite{sealcrypto} to perform the multiplications and additions within the SVM.

\begin{table}[]
    \centering
     \caption{SVM benchmarks used in this work and parameters used in \name. The number of support vectors does not impact latency and energy as they are padded to 4096. For comparison, the accuracy of full-precision SVMs from libSVM \cite{libsvm} with unlimited support vectors is reported. }
    \label{tab:benchmarks}
    \resizebox{0.99\linewidth}{!}{
    \begin{tabular}{|c|c|c|c|c|c|}
         \hline
         Benchmark & Input Bit & Dimension (D) & \# Support & \name\ & LIBsvm  \\
         & Precision & & Vectors & Accuracy & Accuracy \cite{libsvm} \\
         \hline
         MNIST & 3 & 784 & 3841 & 93.85\% & 98.05\% \\
         \hline
         HAR & 3 & 561 & 2466 & 94.64\% & 94.1\%\\
         \hline
         ADULT & 3 & 14 & 596 & 76.00\% & 78.62\%\\
         \hline
    \end{tabular}
    }
   
\end{table}

MTJ based memory technology is already commercially available \cite{everspin,everspin1gb}, however, the devices are expected to significantly improve over the next few years. Hence, we evaluate \name\ with two different MTJs models. A \emph{modern} model, which uses parameters of MTJs which have already been demonstrated, and a \emph{projected} model, which estimates MTJ performance within 3-5 years. These parameters are listed in Table \ref{tab:mtj}. 

\begin{table}[]
    \centering
        \caption{Parameters for MTJ devices.}
    \label{tab:mtj}
      \resizebox{0.8\linewidth}{!}{
    \begin{tabular}{|c|c|c|}
         Parameter & Modern & Projected \\
         $R_P$ & \SI{3.15}{ \kilo \ohm} & \SI{7.34}{ \kilo \ohm} \\
         $R_{AP}$ & \SI{7.34}{ \kilo \ohm} & \SI{76.39}{ \kilo \ohm} \\
         Switching Time & \SI{3}{ \nano \second} \cite{mtjsaida2016sub,3nsmodern} & 1 ns \cite{zabihi2018memory,1nsfuture} \\
         Switching Current & \SI{40}{ \micro \ampere} \cite{mtjsaida2016sub} & \SI{3} {\micro \ampere} \cite{zabihi2018memory} \\
    \end{tabular}
    }
    \vspace{-.4cm}
\end{table}

We model the energy harvester as a constant power source, which fills a \SI{1}{\milli \farad} capacitor (energy buffer) on the chip. In practice, tunable energy buffering systems such as Capybara \cite{colin2018reconfigurable} can be used. The power source increases the voltage on the capacitor up to a specified value, at which point \name\ turns on and consumes energy until the minimum voltage on the capacitor is reached. \name\ then shuts down and waits for the capacitor to recharge. We assume that \name\ has access to sunlight, which can provide a relatively large amount of power for beyond edge devices. We test from \SI{2}{\milli \watt} (0.02 cm$^2$ solar panel \cite{kim2014ambient}) up to \SI{100}{\milli \watt} (1 cm$^2$ solar panel \cite{kim2014ambient}). It is desirable to match the voltage level on the capacitor to the MTJ technology used. Projected MTJs have a lower operating voltage and power draw than modern MTJs. For modern MTJs, the voltage fluctuates between \SI{400}{\milli \volt} and \SI{700}{\milli \volt} and for projected MTJs the voltage fluctuates between \SI{100}{\milli \volt} and \SI{575}{\milli \volt}. As noted in Section \ref{sec:mtjs}, different logic operations require different voltages. We use switched-capacitor converters for upconversion and downconversion to provide all required voltages \cite{SC2jung201423,SC1ramadass2007voltage,SC3harjani2014unified}.

We generate latency and energy estimates of \name\ with an in house simulator which accounts for the overhead due to MTJs (as listed in Tabe \ref{tab:mtj}) and peripheral circuitry, which we extrapolate from NVSIM \cite{dong2012nvsim}. NVSIM gives us the relative share of latency and energy that peripheral circuitry will consume for non-volatile memories with the same size as our computation arrays. We clock \name\ at \SI{30.3}{\mega \hertz} with modern MTJs and at \SI{90.9}{\mega \hertz} with projected MTJs. This gives more than sufficient time for the controller to broadcast the PC and for the drivers to finish logic operations in the memory. This clock rate is conservative, which leaves potential performance improvements on the table. However, since the performance of beyond edge devices will be limited by energy efficiency rather than performance \cite{gobieski2019intelligence,gobieski2019manic} ensuring completion of all instructions far outweighs any potential performance gains.

As discussed in Section \ref{sec:operatingComputation}, the logic operations in the memory are inherently resilient to interruption and we can checkpoint after every instruction with low overhead \cite{resch2020mouse}. However, the reception, transmission, and encoding process do not have this benefit. For transmission and reception, we assume that a power interruption results in the re-transmission or reception of a single element of data. Elements transmitted prior would have been saved in non-volatile memory. For encoding, we assume that the \emph{entire} encoding process must be restarted if interrupted. If interrupted once, \name\ will re-attempt on restart. At this point, the capacitor will be fully charged and \name\ will have sufficient energy to complete the process. For transmission cost, we assume \SI{158}{\pico \joule}/bit \cite{rosenthal2019158} and for encoding we take latency and energy directly from Van der Hagen et. al. \cite{van2021practical} who developed an encoder/encryptor which operates on ciphertexts of the same size as in this work -- \SI{0.3}{\milli \second} and \SI{60}{\micro \joule}, respectively.

\begin{figure}
    \centering
    \includegraphics[width=.45\textwidth]{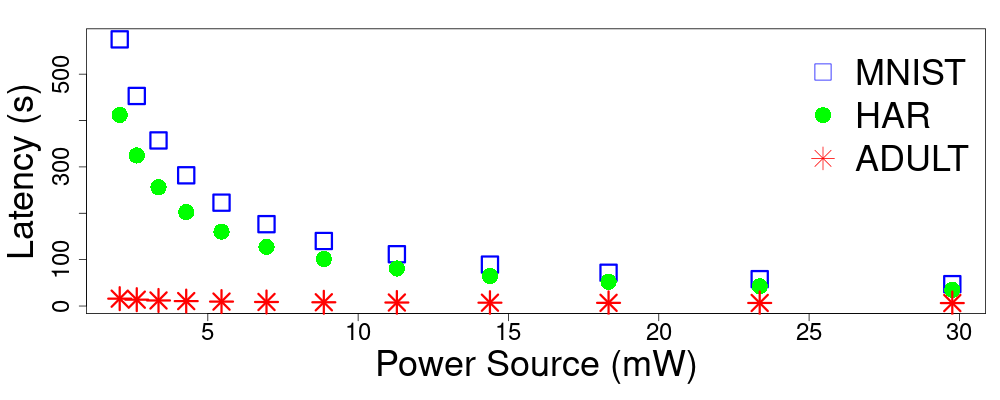}
    \caption{Latency of homomorphic SVM inference with modern MTJs using different power sources. }
    \label{fig:latencyModern}
\end{figure}

\begin{figure}
    \centering
    \includegraphics[width=.45\textwidth]{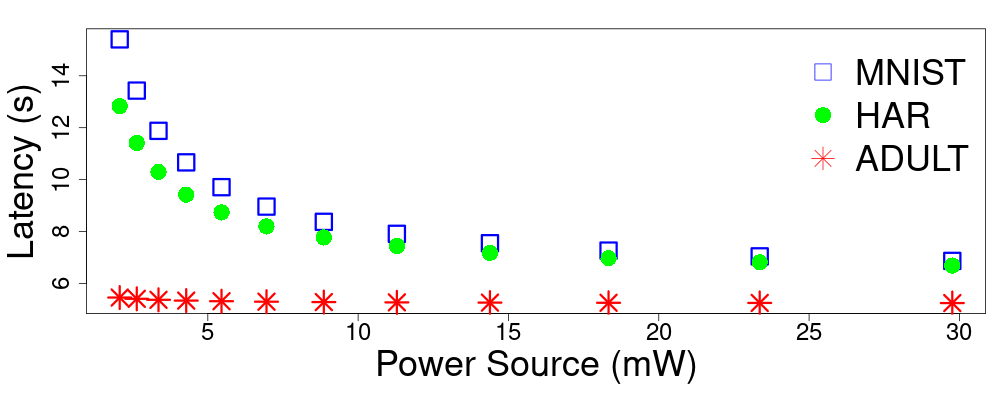}
    \caption{Latency of homomorphic SVM inference with projected MTJs using different power sources. }
    \label{fig:latencyProjected}
    \vspace{-.4cm}
\end{figure}

The latency of one inference (including transmission and encoding) is shown in Figure \ref{fig:latencyModern} for modern MTJs and in Figure \ref{fig:latencyProjected} for projected MTJs. Due to high energy cost of homomorphic computation, the latency is quite high and increases dramatically as the power source reduces. For \name\ to still demonstrate improvement, the latency on \name\ must be less than the latency of local processing on FLY and transmitting to a high-power distance server. For local processing, FLY can perform non-homomorphic inference, since FLY is in a secure location and can use non-encrypted ML models. SONIC \cite{gobieski2019intelligence}, a beyond edge ML accelerator, requires \SI{27}{\milli \joule} to perform MNIST and \SI{12.5}{\milli \joule} to perform HAR. If FLY is operating on thermal energy (due to its deployment within the walls of a building, e.g.), we can assume that its power budget is roughly \SI{60}{\micro \watt} \cite{kim2014ambient,mahan1997thermoelectric}. If FLY is as efficient as SONIC, it will take \SI{450}{\second} to complete MNIST, \SI{208}{\second} to complete HAR, and \SI{8.03}{\second} to complete ADULT. Hence, we can see that \name\ can provide a faster solution than local processing on MNIST/HAR/ADULT if its power budget is \SI{\fasterMNIST}{\milli \watt}/\SI{\fasterHAR}{\milli \watt}/\SI{\fasterADULT}{\milli \watt}, even with modern MTJs. For sending data to a distant server, the cost of transmission is approximately \SI{400}{\micro \joule} per bit. To acquire enough energy to transmit inputs to the server, FLY will take 15,680 seconds for MNIST, 11,220 seconds for HAR, and 280 seconds for ADULT -- much higher than the alternatives. These results are summarized in Table \ref{tab:comparison}.

\begin{table}[]
\caption{As described in Section \ref{sec:problemStatement}, a beyond edge device can (1) Report all results to a remote server, (2) Perform processing locally, or (3) Offload computation to the nearby \name. If FLY has the energy efficiency of the beyond-edge device SONIC \cite{gobieski2019intelligence} and is operating on a reasonable \SI{60}{\micro \watt} (which can be harvested with thermal energy \cite{kim2014ambient,mahan1997thermoelectric}), options (1) and (2) result in the reported latencies. The energy per bit for transmission to the distant server is assumed to be \SI{400}{\micro \joule} \cite{bouguera2018energy}. Also shown is the power required by \name\ to achieve the fastest solution with option (3). Due to being more exposed, \name\ should have access to tens of milliWatts of power \cite{kim2014ambient}. *Results extrapolated.}
    \label{tab:comparison}
    \centering
    \resizebox{.49\textwidth}{!}{
    \begin{tabular}{|c|c|c|c|}
         \hline
         & Option 1: & Option 2:& Option 3:\\
         & Remote Server & Local Processing & \name\ \\
         \hline
         Benchmark & Latency (s) & Latency (s) & Minimum power for \name\  to \\
         & $=$ $E_{FT}/P_F$ & = $E_F/P_F$  & provide fastest solution (mW)  \\
         \hline
         MNIST & 15,680 & 450 & \fasterMNIST \\
         \hline
         HAR & 11,220 & 208 & \fasterHAR \\
         \hline
         ADULT & 280 &  8.03* & \fasterADULT\\
         \hline
    \end{tabular}
    }   
    \vspace{-.4cm}
\end{table}

Consistent with prior work \cite{gobieski2019intelligence,resch2020mouse}, latency is mostly determined by energy efficiency as the device is energy constrained. Consequently, projected MTJs, which are much more energy efficient, enable a significantly reduced latency. Due to the significant overhead of homomorphic computation, despite our lack of bootstrapping, \name\ requires a substantial power budget to remain within reasonable latency constraints. Due to the efficiency of our checkpointing mechanisms, energy consumption is determined mostly by the length of the program, rather than the number of interruptions. The absolute energy consumption of SVM inference on \name\ (evaluated at \SI{2}{\milli \watt}) is listed in Table \ref{tab:energy}. To evaluate the efficiency of \name\ relative to prior work, we compare polynomial multiplication (the core of homomorphic multiplication and the most energy intensive component of our benchmarks) with an X86 CPU, FPGA implementation \cite{nejatollahi2020exploring}, and a processing-in-memory solution, CryptoPim \cite{nejatollahi2020cryptopim}. The comparison is listed in Table \ref{tab:ntt}. As expected, \name\ provides a significant advantage over the CPU and provides a better (yet comparable) efficiency to the FPGA and CryptoPim. It should be noted that both the FPGA and CryptoPim are optimized for \emph{performance}, rather than energy efficiency. However, energy efficiency is the most important metric in the beyond edge domain \cite{gobieski2019manic}, largely because performance is limited by energy efficiency \cite{gobieski2019intelligence}. Additionally, neither the FPGA or CryptoPim have been designed to guarantee correctness during intermittency. Adding tolerance to intermittent operation will come with a performance and efficiency overhead. 

\begin{table}[]
\caption{Energy consumption of homomorphic SVM inference on \name\ for different benchmarks.}
    \label{tab:energy}
    \centering
    \begin{tabular}{|c|c|}
         \hline
         Benchmark & Energy (\SI{}{\micro \joule})  \\
         %\hline
         %SONIC \cite{gobieski2019intelligence} & MNIST & 27,000 \\
         %\hline
         %MOUSE \cite{resch2020mouse} & MNIST & 1,384 \\
         \hline
         MNIST  & \EnergyMNIST\\
         %\hline
         %SONIC \cite{gobieski2019intelligence} & HAR & 12,500 \\
         %\hline
         %MOUSE \cite{resch2020mouse} & HAR & 468 \\
         \hline
         HAR & \EnergyHAR\\
         %\hline
         %SONIC \cite{gobieski2019intelligence} & ADULT & 0.482*\\
         %\hline
         %MOUSE \cite{resch2020mouse} & ADULT & 0.00724 \\
         \hline
         ADULT & \EnergyADULT \\
         \hline
    \end{tabular}
    
\end{table}

\begin{table}[]
\caption{Energy consumption of polynomial multiplication (developed by Nejatollahi et. al \cite{nejatollahi2020exploring}) on \name\ and related work. N is the polynomial size and b is the bitwidth. }
    \label{tab:ntt}
    \centering
    \begin{tabular}{|c|c|}
         \hline
         Architecture (N,b) & Energy (\SI{}{\micro \joule})\\
         \hline
         X86 (1K,16) \cite{nejatollahi2020cryptopim} & 2483.77\\
         X86 (4K,32) \cite{nejatollahi2020cryptopim} & 10864.64 \\
         \hline
         FPGA (1K,16) \cite{nejatollahi2020exploring} & 12.52 \\
         \hline 
         CryptoPIM (1K,16) \cite{nejatollahi2020cryptopim} & 11.04 \\
         CryptoPIM (4K,32) \cite{nejatollahi2020cryptopim} & 178.62 \\
         \hline
         \name\ (1K,16) & 9.68 \\
         \name\ (4K,32) & 54.65 \\
         \hline
    \end{tabular}
    
\end{table}

\begin{figure}
    \subfloat[Modern MTJs]{\includegraphics[width=.24\textwidth]{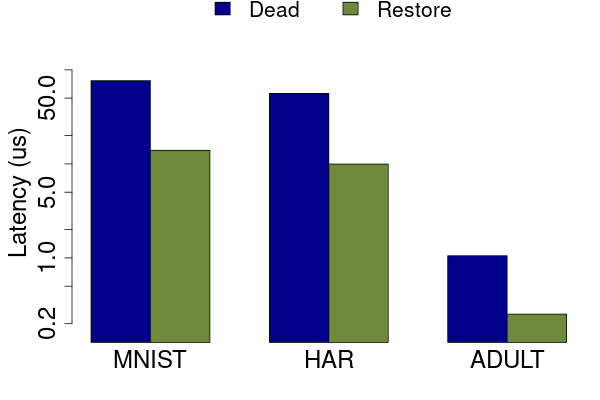}}
    \subfloat[Projected MTJs]{\includegraphics[width=.24\textwidth]{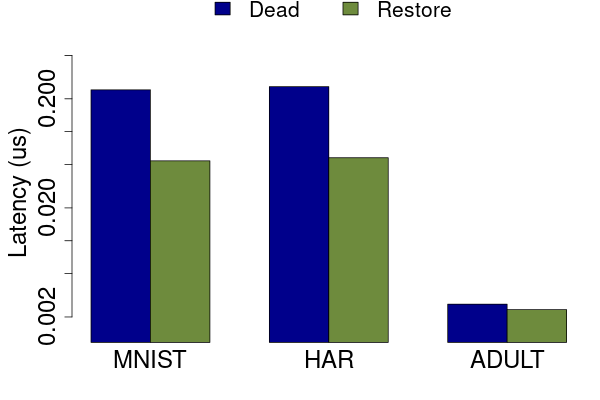}}
    \caption{Latency overhead for correctness during intermittent operation. }
    \label{fig:blatency}
\end{figure}

\begin{figure}
    \subfloat[Modern MTJs]{\includegraphics[width=.24\textwidth]{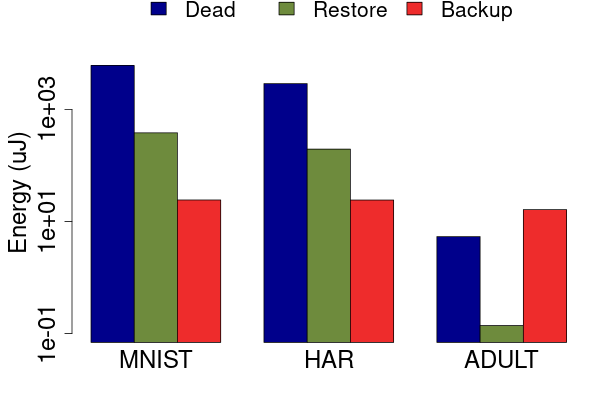}}
    \subfloat[Projected MTJs]{\includegraphics[width=.24\textwidth]{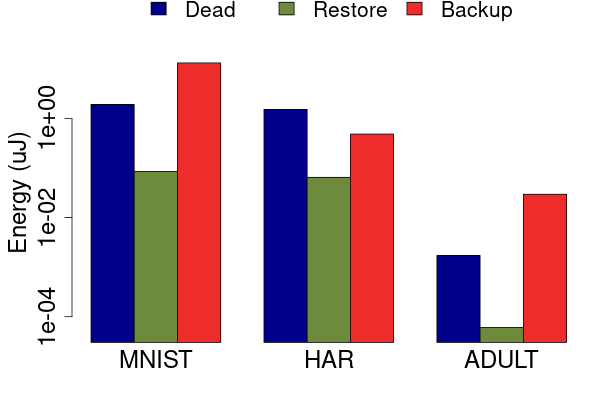}}
    \caption{Energy overhead for correctness during intermittent operation. }
    \label{fig:energy}
    \vspace{-.4cm}
\end{figure}

The impact of intermittent operation can be evaluated by examining the \emph{dead}, \emph{restore}, and \emph{backup} overheads, for both latency and energy \cite{san2018eh}. Dead refers to latency and energy dedicated to re-performing operations after restart which were already performed before shutdown. This accounts for wasted operations -- which completed but the results of which cannot be used. For \name, this comes from re-performing that last in-memory instruction, re-transmitting or receiving a packet of data, and re-encoding input. Restore refers to overhead associated with restarting the device, getting it back into working order after a power outage. For \name, this is the re-activation of rows and columns of the memory arrays. Backup is any operation performed in order to save state prior to shutdown. This typically involves saving the architectural state and storing volatile data to non-volatile memory. As \name\ performs all computation in the memory, data backup occurs automatically. Hence, the only backup cost for \name\ is saving the architectural state, setting the SR, PC, and parity bit. The latency overhead for each of these is shown in Figure \ref{fig:blatency} and the energy overhead is shown in Figure \ref{fig:energy}. The shown overheads are evaluated at \SI{20}{\milli \watt}, the lowest power considered, which results in the maximum number of restarts, and therefore, the maximum overhead. 

Overheads for projected MTJs are less because \name\ will have to restart less often due to their greater energy efficiency. Dead energy involves re-performing instructions (which typically involve performing many parallel logic operations). %For this reason, 
Overall, projected MTJs are much more efficient than modern MTJs. Dead energy is \STTmED \% of the total for modern MTJs and \STTfED \% for projected MTJs. The restore energy is also much lower for projected MTJs, as it is also incurred by only restarts where the peripheral circuitry needs to be reset. Restore energy is \STTmER \% of the total for modern MTJs and \STTfER \% of the total for projected MTJs. As it involves only updates to architectural state variables, the backup energy tends to be much lower. The backup energy is \STTmEB \% of the total for modern MTJs and \STTfEB \% of the total for projected MTJs. Backup energy is incurred on every instruction, so it is not largely impacted by the number of restarts. 

Backup has no associated additional latency as it occurs during the execution of each instruction. The dead and restore latency are functions of the number of restarts. Dead (restore) latency is \STTmLD \% (\STTmLR \%) of the total compute latency for modern MTJs. For projected MTJs the dead (restore) latency is a negligible \STTfLD \% (\STTfLR \%).

%% file: sections/limitations.tex
\vspace{-.4cm}
\section{Limitations and Potential Improvements}
\label{sec:limitations}
In this section we discuss the limitations of \name\ and potential improvements. While relevant, these extensions are beyond the scope of this paper or are left for future work.
\vspace{-.2cm}
\subsection{Communication Protocol}
In this work we use a simplified communication scheme, where FLY and \name\ transfer data back and forth. There are a number of complicating factors which can disrupt proper operation. FLY may begin transmission but then loose power for an extended period of time. In this case, \name\ would remain in reception mode, spending its entire time waiting for input that does not come. \name\ should contain some method of cancelling a reception and revert to an idle and available mode. While a variant of watchdog timer can be deployed to this end, data transfers during intermittent operation pose a considerable challenge. Additionally, if \name\ goes without power for an extended period of time, it may work on input which is old and no longer relevant. Keeping beyond edge devices working on relevant tasks across time is an important problem \cite{surbatovich2021automatically}.
\vspace{-.2cm}
\subsection{Communication Energy and Deployment}
Communication between FLY and \name\ will only be efficient if they can be deployed in close proximity. Bluetooth Low Energy (BLE) only allows for communication within tens of meters. At further distances, they must resort to long-range (LoRa) technology \cite{bouguera2018energy}, which can undo the benefit \name\ provides. 
\vspace{-.2cm}
\subsection{Power Limitations}
Homomorphic computation, even without bootstrapping, remains an energy costly process. Even with the efficient computation enabled by \name, collecting enough energy beyond edge remains a challenge. Even simple computations will take a long time, which may be insufficient for a number of applications. Further methods to increase energy efficiency may be necessary for practical implementation. A possible solution is implementing \emph{fleets} of beyond edge devices \cite{denby2019orbital,denby2020orbital}. As such, each individual device could compute slowly, but a high throughput could be achieved with their combined effort. 
\vspace{-.2cm}

%% file: sections/related.tex
\section{Related Work}
\label{sec:related}
Power delivery is particularly challenging for beyond edge devices. Tunable systems like Capybara \cite{colin2018reconfigurable} have been designed to optimally store and deliver energy. We assume \name\ relies on such a system.

Many low power processors and ML accelerators exist \cite{seok2008phoenix,pudiannao,conti2018xnor,wang201928,jia2019programmable,zhang2019memory,valavi201964}, however, these architectures do not guarantee correctness during intermittent operation. Adding such guarantees adds a large performance and efficiency overhead. Numerous beyond edge devices have been designed to function correctly in intermittent contexts \cite{liu2015ambient,ma2015architecture}, including ML accelerators \cite{gobieski2019intelligence}. These devices have more traditional architectures, which have been augmented with nearby non-volatile memory for fast backup operations. Much research has been dedicated to tolerating intermittency for more general purpose architectures, including lowering checkpointing overheads and clever methods of detecting power outages \cite{cleancutcolin2018termination,coatiruppel2019transactional,ganesan2019s,mementosransford2011mementos,chaincolin2016chain,alpacamaeng2017alpaca,mayflyhester2017timely,ratchetvan2016intermittent,clankhicks2017clank,hibernusbalsamo2016hibernus++,jayakumar2014quickrecall,aouda2014incremental,balsamo2016graceful,berthou2017peripheral,liu2016lightweight,lukosevicius2017using} along with modifications of software \cite{chinchillamaeng2018adaptive,dinolucia2015simpler}. \name's advantage over these designs is significantly more energy efficient in-memory logic and simplified  checkpointing mechanisms.

Resch et. al. \cite{resch2020mouse} have proposed an in-memory intermittent safe accelerator. However, the architecture maintains standard memory format, which adds unnecessary space and energy overhead, and is capable only of 1D computations (only along rows or only along columns).  In contrast, \name's architecture is capable of 2D computations (similar to a crossbar, along rows and columns, but digital) and is tightly tailored to homomorphic SVM inference. Mapping homomorphic inference onto \cite{resch2020mouse} would incur a large communication (read/write) overhead due to the required intra-array data movement. ResiRCA \cite{qiu2020resirca} uses in-memory computation to accelerate parts of ML inference and adapts the amount of parallelism to match the amount of harvested energy. However, they rely on a battery to maintain a controlling CPU.

Significantly, none of the prior work mentioned thus far has accelerated homomorphic inference, which is necessary to maintain security in un-trusted environments. To the best of the author's knowledge, no beyond edge homomorphic accelerator exists. However, much prior work has been done on accelerating homomorphic computation with continuous power. This includes software optimization \cite{reagen2021cheetah}, FPGA implementations \cite{nejatollahi2020exploring,turan2020heaws} and processing-in-memory implementations like CryptoPim \cite{nejatollahi2020cryptopim}. CryptoPim developed strategies to minimize latency and energy overhead for NTT operations, the core of homomorphic multiplication. While CryptoPim is not designed to handle intermittent operation, we do use strategies they developed, such as efficient modulus operations, in our work.

Efficient encoding and encryption is required to make homomorphic computation feasible in edge or beyond edge domains. Van Der Hagen et. al. \cite{van2021practical} developed energy efficient accelerators to this end.

%% file: sections/conclusion.tex
\section{Conclusion}
\label{sec:conclusion}
Beyond edge devices operate on harvested energy, significantly increasing deployment capabilities and widening the potential applications. However, these devices 
are by construction extremely
%can be significantly 
limited by the power available. In this paper we propose a beyond edge, intermittent safe accelerator which can compute homomorphically, and which can be used as a local server to offload computation. The deployment capability of this accelerator can significantly reduce the communication cost for other beyond edge devices to offload computation. The accelerator is capable of maintaining security by performing homomorphic computation and its unique architecture enables this to be done within a reasonable power budget.